\documentclass[reprint,superscriptaddress,nofootinbib,amsmath,amssymb,aps,showkeys,showpacs,prd]{revtex4-2}

\usepackage{caption}
\usepackage{csquotes}
\usepackage{orcidlink}
\usepackage{footmisc}
\usepackage{extpfeil}
\usepackage{subcaption}
\usepackage{graphicx}
\usepackage{dcolumn}
\usepackage{bm}
\usepackage{mathrsfs}
\usepackage{graphicx}
\usepackage{braket}
\usepackage{enumitem}
\usepackage[dvipsnames]{xcolor}
\usepackage{physics}
\usepackage{color}
\usepackage{cleveref}
\usepackage{siunitx}
\usepackage{xcolor}
\usepackage{soul}
\usepackage[T1]{fontenc} 
\usepackage{hyperref}
\hypersetup{
	colorlinks=true,
	linkcolor=blue,
	citecolor=red}
\usepackage{tikz}
\usetikzlibrary{positioning,calc,arrows.meta}
\usetikzlibrary{shapes.geometric, arrows, positioning}

\tikzstyle{graybox} = [
draw=black!50,
fill=black!3,
rounded corners=4pt,
inner sep=7pt
]
\tikzstyle{box} = [
rectangle, 
rounded corners,
minimum width=1.75cm, 
minimum height=1cm,
text centered, 
draw=black, 
fill=gray!20
]

\tikzstyle{arrowG} = [->, >=stealth, draw=green!70!black, line width=2.5pt]  
\tikzstyle{arrowY} = [->, >=stealth, draw=yellow!60!black, line width=2.5pt]   

\begin{document}
	
	\title{Subtleties in non-equilibrium horizon thermodynamics of modified gravity theories}
	
	\author{Vishnu A Pai \orcidlink{0000-0003-4161-3383}
	}
	\email{vishnuajithj@gmail.com, vishnuajithj@cusat.ac.in}
	\affiliation{Department of Physics, Cochin University of Science and Technology, Cochin -682022, India}
	
	\author{Vishnu S Namboothiri \orcidlink{0000-0001-7740-9179}
	}
	\email{ramharisindu@gmail.com}
	\affiliation{Department of Physics, Cochin University of Science and Technology, Cochin -682022, India}

	\author{Titus K Mathew \orcidlink{0000-0002-9307-3772}
	}
	\email{titus@cusat.ac.in}
	\affiliation{Department of Physics, Cochin University of Science and Technology, Cochin -682022, India}
	
	
\begin{abstract}
	Thermodynamic interpretations of gravity often arise from applying the Clausius relation to spacetime horizons. In modified gravity theories with higher-order equations of motion, such as $f(R)$ and scalar-tensor gravity, this relation generally acquires additional entropy-production term. In this context, two distinct formulations have been proposed in literature: the non-equilibrium approach of Eling, Guedens, and Jacobson based on local Rindler horizons, and the thermodynamic formulation of cosmological apparent horizons in FLRW spacetimes. In this article, we present a detailed analysis of these approaches, and show that, even though both employ identical entropy balance relations that resemble non-equilibrium thermodynamics, the exact origin and role of each entropy-production term is fundamentally different. In the Rindler-horizon framework the extra term follows directly from consistency requirements related to the Bianchi identity, whereas in the apparent-horizon approach it is introduced solely to recover the Friedmann equations.  Furthermore, we will see that the latter non-equilibrium contribution enters directly into dynamical equations of gravity, while the former does not. Finally, we also highlight the fact that thermodynamic descriptions of horizons in such modified gravity are not unique, and that equilibrium, and non-equilibrium descriptions can arise from different choices of thermodynamic variables. A clear understanding of these distinctions is therefore crucial for establishing a consistent and physically meaningful thermodynamic foundation for gravity beyond general relativity.
\end{abstract}
\maketitle

\section{Introduction}

The intriguing connection between gravitation and thermodynamics has long suggested that the dynamics of spacetime may in fact be governed by thermodynamic principles encoded with horizon surfaces. A precise mathematical formulation of this idea was first given by Jacobson in his seminal work \cite{Jacobson:1995ab}, where the Einstein field equations were derived from the Clausius relation applied to local Rindler horizons. In this approach, the entropy of the horizon is assumed to be proportional to its area, and the energy flux across the horizon is interpreted as heat. Under these assumptions, the Einstein equations emerge as an equation of state of the system, indicating that the dynamics of spacetime may arise from underlying thermodynamic principles. If this thermodynamic origin of gravity is universal, it is natural to expect that the field equations of more general gravity theories should also emerge from similar thermodynamic arguments.

Motivated by this idea, Eling, Guedens, and Jacobson (EGJ) later investigated whether the same reasoning can be extended to modified gravity theories in which the horizon entropy density deviates from the standard area law. In particular, they considered the case where  the entropy density of the Rindler horizon becomes an arbitrary function of the Ricci scalar \cite{Eling:2006aw}. They found that in such cases the equilibrium Clausius relation is insufficient to reproduce the correct gravitational field equations. Instead, the thermodynamic description of the system becomes intrinsically non-equilibrium and requires the inclusion of an entropy production term for satisfying local energy conservation. The resulting non-equilibrium entropy balance relation takes the form
\begin{equation}\label{EGJ}
	dS = \delta Q/T + d_i S,
\end{equation}
where $d_i S$ represents the entropy production term \cite{Eling:2006aw,Chirco:2009dc}, $S$ is the horizon entropy, $T$ is the temperature, and $\delta Q$ is the heat flux crossing the horizon. In Einstein gravity the equilibrium situation is recovered, for which $d_i S = 0$.

A related but distinct thermodynamic approach has also been proposed in cosmological spacetime, where the dynamical equations of the Universe are derived from thermodynamic relations associated with the apparent-horizon (which is same as the Hubble horizon for a spatially flat Universe). Unlike black-hole spacetimes which have event horizons, Friedmann-Lema\^{i}tre-Robertson-Walker (FLRW) universes are generally time dependent. The event horizon, when it exists, depends on the entire future evolution of the spacetime and is therefore not suitable for defining local thermodynamic relations \cite{Wang:2005pk}. However, several studies have shown that the Clausius relation can hold on the apparent horizon even though it fails on the event horizon \cite{Cai:2005ra,Akbar:2006kj,Tian:2014sca}. For this very reason, the apparent horizon is widely regarded as the natural thermodynamic boundary in cosmological spacetime. Within this framework, numerous articles have shown that Friedmann equations in Einstein gravity can be expressed in the form of Clausius relation projected on to the apparent horizon of the Universe by taking Wald-entropy as the entropy of the horizon \cite{Cai:2005ra,Akbar:2006kj,Tian:2014sca}. In this picture, the cosmic expansion can be viewed as a thermodynamic process transpiring in the horizon-bulk system.

This framework has also been extended to modified gravity theories, including $f(R)$ gravity, scalar-tensor models, and braneworld scenarios \cite{Cai:2006rs,PhysRevD.90.104042,Gong:2007md,Bamba_2010,BAMBA2010101,GE200749}. In such theories the horizon entropy generally differs from the standard Bekenstein-Hawking form because of additional gravitational degrees of freedom. As a result, reproducing the modified Friedmann equations from thermodynamic arguments typically requires one of two approaches. One possibility is to introduce an additional entropy production term in the Clausius relation in an ad hoc manner, leading to the modified entropy balance law \cite{Cai:2006rs,PhysRevD.90.104042}
\begin{equation}\label{CAH}
	dS + dS_p = \delta Q/T.
\end{equation}
Alternatively, one may suitably define a Misner-Sharp masslike function for the Horizon which will redefine the energy flux crossing the apparent horizon, so that the standard equilibrium Clausius relation remains valid and no additional entropy production term is required \cite{Gong:2007md,Bamba_2010,BAMBA2010101}. In other words, one has the freedom to adopt either equilibrium or non-equilibrium descriptions while investigating thermodynamic aspects of cosmological apparent-horizon (CAH), without altering the actual dynamical equations of gravity.

This freedom introduces an ambiguity, since it becomes unclear whether the apparent horizon in these gravity theories with higher-order equations of motion obey non-equilibrium or equilibrium thermodynamics, a detailed discussion is provided in Sec. \ref{eqnoneq}. Moreover, in Sec. \ref{MFRWE} and \ref{comparison}, we show that unless one provides a well-defined theoretical prescription for the entropy production term, attempts to derive the modified Friedmann equations using non-equilibrium approach within CAH framework can lead to circular arguments.

Although both EGJ's and CAH analysis may appear similar, due to the analogous form of Clausius relations, they essentially arise from two fundamentally different physical setups, and hence represents distinct thermodynamic formulations of gravitational systems. Understanding the differences between these two approaches is therefore important for properly interpreting thermodynamic descriptions of gravity in modified theories and cosmological settings. The main goal of this article is to provide a detailed comparison of these two frameworks, emphasizing the different physical origins of their entropy production terms and their roles in the thermodynamic description of an expanding universe.

\section{EGJ framework: Deriving modified field equations from the non-equilibrium Clausius relation}\label{MFE}

In this section we will go-through the results derived by Eling, Gudenson and Jacobson (EGJ) in \cite{Eling:2006aw}, where they derive modified field equations in $f(R)$ gravity from the non-equilibrium Clausius relation by applying it on a local Rindler horizon patch whose entropy density is the same as the Wald entropy in $f(R)$ gravity. We will focus on highlighting the important points which sets up the premise for the discussions in Sec. \ref{comparison}.

In f(R) gravity, the Wald entropy associated with causal horizon is of the form,
\begin{equation}
	S = \frac{1}{4G} \int_{\mathcal{H}} f'(R)\, dA,
\end{equation}
where, $f'(R)=df(R)/dR$. Varying this relation along the null generators parameterized by the affine parameter `$\lambda$', we get,
\begin{equation}
	dS =\frac{1}{4G}\int \left[f' \frac{d(dA)}{d \lambda}+dA\frac{df'}{d\lambda}\right] d\lambda
\end{equation}
Using the equation for expansion of the congruence of null geodesics generating the horizon, which is given as, $\theta=d\{\ln\left(dA\right)\}/d\lambda$, and denoting derivative with respect to $\lambda$ as an over dot, one gets the result,
\begin{equation}\label{dS}
	dS =\frac{1}{4G}\int \left[\theta f' + \dot{f}'\right] d\lambda dA
\end{equation}
The heat flux, defined as the mean flux of boost energy current crossing the horizon, is given as;
\begin{equation}
	\delta Q=\int T_{ab}\;\chi^a \,d\Sigma^b
\end{equation}
Here, $\chi^a$ represents the boost Killing vector field, and in flat space-time, it is related to affine parameter via the relation, $\chi^a=-\lambda k^a$, where $k^a$ represents the Horizon tangent vector, $k^a=dx^a/d\lambda$. Note that temperature of causal Horizon is defined as, $T=\hbar/2\pi$. Using this fact, one obtains,
\begin{equation}\label{dQ/T}
	\frac{\delta Q}{T}=\frac{2\pi}{\hbar}\int T_{ab}\;k^a\,k^b \left(-\lambda\right) d\lambda\,dA
\end{equation}
If one now imposes the Clausius relation, $dS=dQ/T$ then, then it is clearly visible that; if expansion rate $\theta$ vanishes at the point $p$ on the causal Horizon, then integrand of equation (\ref{dS}) is non-zero as it has an additional term $\dot{f'}$. This cannot match with the $\lambda^{th}$ order term in the energy flux equation (\ref{dQ/T}). Hence, one imposes the constraint that, instead of $\theta$ vanishing, one mush have, 
\begin{equation}\label{cons}
	\left[\theta f' + \dot{f}'\right]\Big|_{p}=0
\end{equation}
Then, if one retains only the $\lambda^{th}$ order term in equation (\ref{dS}), use the Raychaudhuri equation (with vanishing shear) given as,
\begin{equation}\label{Ray}
	\frac{d\theta}{d\lambda} = -\frac{1}{2}\theta^2 - R_{a b}\, k^a k^b
\end{equation}
along with the geodesic equation, $k^a k^b_{\,\;;a}=0$ , one arrives at the expression,
\begin{equation}\label{dS4}
	dS=\int \left[f'_{,\,a\,b}-R_{a b}f' - \frac{3}{2f'}\,f'_{,a}\,f'_{,b}\right] \frac{\lambda }{4G}\,k^a k^b\,d\lambda dA
\end{equation}
Equating this with the flux relation obtained in equation (\ref{dQ/T}), one can see that integrand of both equations must match for all null vectors $k^a$. This leads to the equation,
\begin{equation}\label{E0}
	R_{a b}\,f' + \frac{3}{2f'}\,f'_{,a}\,f'_{,b} - f'_{,\,a\,b} +\Psi g_{ab}= \frac{\kappa}{\hbar}T_{ab}
\end{equation}
where, $\Psi$ is an undetermined scalar function and we have denoted $\kappa=8\pi G$. 

To determine this function, we can consider the fact that energy momentum tensor is divergence free and hence set, $T_{ab}^{\;\;\;;a}=0$. This means, the left hand side (LHS) of equation (\ref{E0}) must also be divergence free. This gives the constraint,
\begin{equation}
	\left[R_{a b}\,f' + \frac{3}{2f'}\,f'_{,a}\,f'_{,b} - f'_{,\,a\,b} +\Psi g_{ab}\right]^{;a}=0
\end{equation}
Using the distributive property of covariant derivatives, we can simplify the above expression as,
\begin{equation}
	\left[R_{a b}\,f' - f'_{,\,a\,b} \right]^{;a} + \left[\frac{3}{2f'}\,f'_{,a}\,f'_{,b}\right]^{;a} +\Psi^{;a} g_{ab}=0
\end{equation}
However, using the Lagrangian density, $\mathcal{L}=f(R)$, one can show that, 
\begin{equation}
	\left[R_{a b}\,f' - f'_{,\,a\,b} \right]^{;a} =\left(\mathcal{L}/2 -\Box f\right)_{,\,b}
\end{equation}
Furthermore, lowing the index on the third term and replacing covariant derivative with a partial derivative (since $\Psi$ is a scalar function), we get,
\begin{equation}
	-\left[\mathcal{L}/2 - \Box f+\Psi\right]_{,\,b}=\left[\frac{3}{2f'}\,f'_{,a}\,f'_{,b}\right]^{;a}\equiv\Theta_{,\,b}
\end{equation}
Note that, since LHS in the above equation is a gradient of a scalar and RHS represents a covariant derivative, the above relation cannot be generally true. This is because, in general, the quantity in the RHS cannot be written as the gradient of a scalar. Hence the $\Theta$ term arising in the above equation is problematic, as it acts as hindrance in obtaining a covariant field equation that respects local matter conservation law. To overcome this issue, one can modify the Clausius relation by adding an entropy balance term (non-equilibrium entropy production) $d_i S$ and choose its functional form such that the unwanted term gets canceled out. This requirement uniquely specifies the functional form of the entropy production as,
\begin{equation}\label{diS}
	d_i S=\int \frac{3}{2f'}\,f'_{,a}\,f'_{,b} \, k^a\,k^b \left(-\lambda\right)d\lambda dA
\end{equation}
Subsequently, one obtains the standard modified field equation in $f(R)$ gravity, 
\begin{equation}
	f' R_{ab} - \nabla_a\nabla_b f'+ \left[\,\Box f' - f/2\,\right] g_{ab}= \kappa\,T_{ab}.
\end{equation}
Note that if one considers the above definition of entropy production, then it must be added on RHS of Clausius relation such that it automatically cancels the analogous term appearing on geometry side. Hence, in this case the Clausius relation is modified as, $dS=dQ/T +d_i S$. One can also see that the entire purpose of the $d_iS$ term is to cancel certain spurious terms arising when horizon entropy with additional degree of freedom is varied. The remaining terms are what forms the correct gravitational field equations. Therefore, the entropy production term $d_iS$ does not enter the dynamical equations directly.

\section{CAH framework: Deriving modified Friedmann equations from the non-equilibrium Clausius relation}\label{MFRWE}

Let us now look at the alternative approach where the Friedmann equations, and not the Einstein equations, are derived from the Clausius relation applied to apparent-horizon of the Universe. In most cases authors adopt an inverse approach where Friedmann equations of some particular gravity theory is re-expressed as a first-law like relation \cite{PhysRevD.75.084003}. However, if thermodynamic formulation is considered as fundamental, then one must be able to do the vice-versa as well, i.e. derive dynamical equations by imposing the first law of thermodynamics to apparent-horizon.

We define all thermodynamic quantities associated with the apparent horizon of the Universe by following Hayward's original definitions \cite{Hayward:1997jp,Hawking:1976de}. Cosmological apparent horizon (CAH) is defined quasi-locally from the vanishing of the expansion of radial null congruences, and therefore represents an instantaneous causal boundary of the spacetime. Geometrically, it is a spatial two-surface with radius $r_A$ and surface area $A$ given as
\begin{equation}
	r_A = \frac{1}{\sqrt{H^2 + k/a^2}}\quad\textbf{;}\quad A = 4\pi r_A^2 .
\end{equation}

As per the original definition, the energy flux crossing the apparent horizon of the Universe during an infinitesimal time interval $dt$, during which the horizon is assumed to be approximately static, is given by $dE_f=-A\psi_t$, where $\psi_t$ represents the energy flux density in the static horizon limit. For a perfect fluid with energy density $\rho$ and pressure $p$, this quantity is defined as $\psi_t=-(\rho+p)Hr_{A}dt$. This immediately implies
\begin{equation}
	-dE_f= -A\psi_t=A\left(\rho+p\right)Hr_{A}dt .
\end{equation}

One then considers the entropy associated with the apparent horizon in $f(R)$ gravity to be the Wald entropy given by $S_{H}=Af'/4G$, while the temperature is taken to be the Gibbons-Hawking temperature $T_H=1/2\pi r_A$ \cite{Akbar:2006kj}. Substituting these relations into the standard Clausius relation, $TdS=\delta Q=-dE_f$, one obtains
\begin{equation}\label{sad}
	r_A\dot{f'}+ 2\, \dot{r}_A f'= \kappa r_A^3 H\left(\rho+p\right) .
\end{equation}
Substituting the definition of $r_A$, we obtain
\begin{equation}
	\dot{r}_A=-Hr_A^3\left(\dot{H}-k/a^2\right) ,
\end{equation}
which can then be substituted in Eqn.~\eqref{sad} to obtain
\begin{equation}\label{Fried}
	2f'\dot{H}=-\kappa\left(\rho+p\right)+H\dot{f'}+\frac{k\dot{f'}}{a^2 H}+2f'(k/a^2) .
\end{equation}
Even though this equation might resemble the Friedmann equations in $f(R)$ gravity, an exact comparison with the actual ones obtained by extremizing the action, as given in \ref{fRaction}, reveals that certain essential terms are missing (most notably the term $-\ddot{f^{\prime}}$), while some spurious terms are present (such as $k\dot{f^{\prime}}/a^2H$) in Eq. \eqref{Fried}. Hence, to obtain the correct dynamical equations, it is necessary to modify the initial Clausius relation in some way. Two alternative approaches have been proposed in literature;

(a) Inclusion of an additional entropy production term so that the spurious terms are canceled and the essential terms are incorporated \cite{Cai:2006rs,PhysRevD.90.104042}. In such cases, some times the energy fluxes are also modified, and based on the choice of energy flux, the corresponding function form of entropy-production required to retain the correct equations of motion changes accordingly.

(b) Redefining the Misner-Sharp mass by proposing suitable masslike function and modifying the energy flux through the apparent horizon \cite{Gong:2007md,Bamba_2010,BAMBA2010101}, in which case the equilibrium Clausius relation is retained.

A comparison between Eqn.~\eqref{Fried} and \eqref{eq:friedmann2} reveals that the entropy-production term required for obtaining the exact Friedmann equation in $f(R)$ gravity from the non-equilibrium relation $dS+dS_p=\delta Q/T$ should be,
\begin{equation}
	T\frac{dS_p}{dt}= -\frac{Hr_A^3}{2G}\left[\ddot{f}^{\prime} + \frac{k}{a^2 H} \dot{f}^{\prime}\right]
\end{equation}
Indeed, this entropy production term is entirely different from the one appearing in the EGJ approach, i.e. Eqn.~\eqref{diS}. We also note that the contribution arising from $dS_p$ enters directly into the Friedmann equations and can therefore influence the cosmological dynamics. 

\section{Detailed analysis of the two non-equilibrium approaches and associated entropy production terms}\label{comparison}

We will now compare the two approaches discussed in the previous sections in detail for better understanding the thermodynamic aspects of gravitational dynamics in cosmological spacetimes. 

We begin with the standard equilibrium case. In Einstein gravity it is well known that the equilibrium Clausius relation, $dS=\delta Q/T$ is sufficient to reproduce the correct dynamical equations in both of these two seemingly different contexts. First, in Jacobson’s local-horizon construction \cite{Jacobson:1995ab}, applying the Clausius relation to local Rindler horizon patches directly leads to the Einstein field equations. Second, in cosmology, applying a similar thermodynamic relation to the apparent horizon of an FLRW universe reproduces the correct Friedmann equations governing cosmic expansion \cite{Cai:2005ra,Tian:2014sca}. 

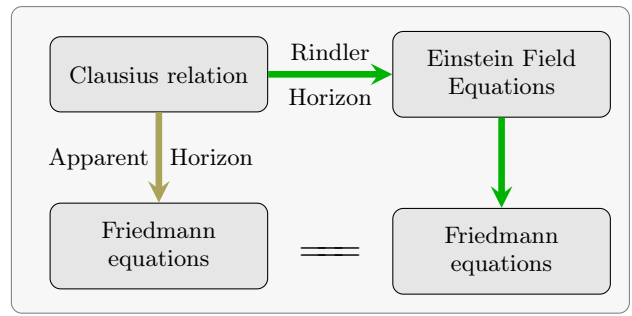
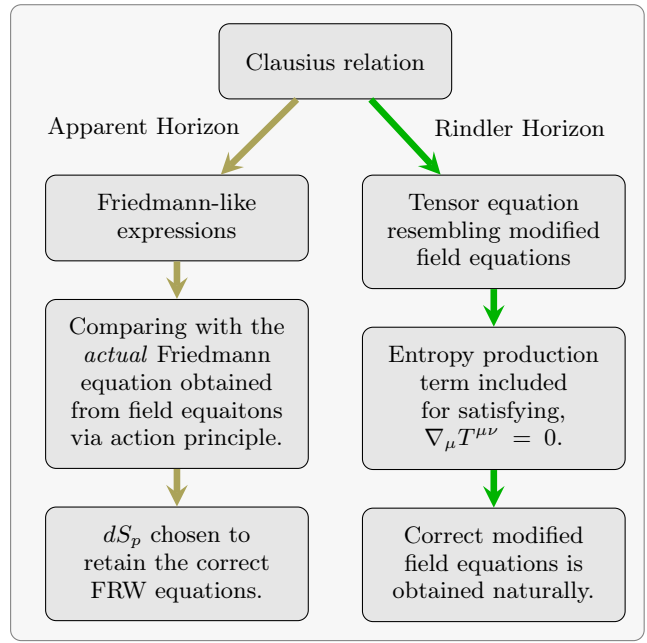
\begin{figure}
	\centering
	\begin{subfigure}{\columnwidth}
		\centering
		\begin{tikzpicture}
			\node[graybox] {
				\begin{tikzpicture}[node distance=2cm]
					\node (cl1) [box, text width=2.4cm] {Clausius relation};
					\node (efeq) [box, right=1.65cm of cl1, text width=2.4cm] {Einstein Field\\Equations};
					\node (fr1)  [box, below=1.2cm of efeq, text width=2.4cm] {Friedmann equations};
					\node at ($(cl1)!0.5!(efeq) + (0,0.3)$){Rindler};
					\node at ($(cl1)!0.5!(efeq) + (0,-0.3)$){Horizon};
					\node (fr2) [box, below=1.2cm of cl1, text width=2.4cm] {Friedmann equations};
					\node at ($(cl1)!0.70!(fr2) + (-0.8,0.47)$) {Apparent};
					\node at ($(cl1)!0.70!(fr2) + (0.7,0.5)$){Horizon};
					\draw[arrowG] (cl1) -- (efeq);
					\draw[arrowG] (efeq) -- (fr1);
					\draw[arrowY] (cl1) -- (fr2);
					\node at ($(fr1)!0.5!(fr2)$) {\Large $\mathbf{\xlongequal{\quad}}$};
				\end{tikzpicture}
			};
		\end{tikzpicture}
		\caption{Equilibrium horizon thermodynamics}\label{4DPLna}
	\end{subfigure}
	\vspace{0.5cm}
	\begin{subfigure}{\columnwidth}
		\centering
		\begin{tikzpicture}
			\node[graybox] {
				\begin{tikzpicture}[node distance=2cm]
					\node (cl1) [box, text width=2.6cm] {Clausius relation};
					
					\node (tnsr) [box, below right=1.0cm and -1.2cm of cl1, text width=3cm]
					{Tensor equation resembling modified field equations};
					
					\node (exp) [box, below left=1.0cm and -1.2cm of cl1, text width=3cm]
					{Friedmann-like expressions};
					
					\node (fr1) [box, below=0.5cm of tnsr, text width=3cm]
					{Entropy production term included for satisfying, $\nabla_\mu T^{\mu\nu}=0$.};
					
					\node (DSP) [box, below=0.5cm of exp, text width=3cm]
					{Comparing with the \emph{actual} Friedmann equation obtained from field equaitons via action principle.};
					
					\node (fr2) [box, below=0.5cm of DSP, text width=3cm]
					{$dS_p$ chosen to retain the correct FRW equations.};
					
					\node (efeq)  [box, below=0.5cm of fr1, text width=3cm]
					{Correct modified field equations is obtained naturally.};
					
					\node at ($(cl1)!0.5!(tnsr) + (1.4,0.25)$) {Rindler Horizon};
					\node at ($(cl1)!0.5!(tnsr) + (-3.6,0.25)$) {Apparent Horizon};
					
					\draw[arrowG] (cl1) -- (tnsr);
					\draw[arrowY] (cl1) -- (exp);
					\draw[arrowY] (exp) -- (DSP);
					\draw[arrowY] (DSP) -- (fr2);
					\draw[arrowG] (tnsr) -- (fr1);
					\draw[arrowG] (fr1) -- (efeq);
				\end{tikzpicture}
			};
		\end{tikzpicture}
		\caption{Non-equilibrium horizon thermodynamics.}\label{4DPLnb}
	\end{subfigure}	
	\caption{Flowcharts describing the standard approach employed for deriving the Friedmann equations, and Einstein equation, from thermodynamic relations.}
	\label{4DPLn}
\end{figure}

These two derivations are conceptually different: the former is a local construction based on infinitesimal Rindler horizons in an arbitrary spacetime, while the latter relies on the global symmetries of FLRW geometry. Nevertheless, both approaches yield mutually consistent results when the underlying theory is general relativity. This equivalence is illustrated schematically in Fig. \ref{4DPLna}. Starting from the equilibrium Clausius relation, the local Rindler-horizon construction leads to the Einstein field equations, which in turn reduce to Friedmann equations in a homogeneous and isotropic spacetime. At the same time, an independent application of the Clausius relation to the cosmological apparent horizon leads directly to the same Friedmann equations. 

Similar agreement can be seen in Lovelock gravity theories, where the entropy of the apparent-horizon is the corresponding Wald entropy. In these contexts, the equilibrium thermodynamic relation remains sufficient to recover the corresponding gravitational  field equations \cite{Cai:2009qf}. This intriguing consistency strongly supports the view that general relativity admits a genuine equilibrium thermodynamic interpretation.

The situation changes when one considers modified theories of gravity with additional degree of freedom, such as $f(R)$ or scalar-tensor gravity theories. Due to the presence of additional dynamical degrees of freedom, such theories are generally associated with  field equations that lead to higher-order equations of motion. Consequently, if one considers usual definitions for horizon entropy and heat flux, then simple equilibrium Clausius relation is found to be no longer sufficient to reproduce the correct gravitational dynamics in both frameworks \cite{Eling:2006aw,Cai:2006rs,PhysRevD.90.104042}. The resulting logical structure is illustrated in Fig. \ref{4DPLnb}, where the equilibrium relation leads only to some intermediate expression that do not coincide with actual dynamical equations.

Within the framework proposed by EGJ, the resolution of this problem requires a generalization of the Clausius relation to a non-equilibrium entropy balance relation, $dS=\delta Q/T+d_iS$. Here the additional term $d_iS$ represents internal entropy production. As shown in Sec.~\ref{MFE}, this term arises naturally when the entropy density of the horizon depends on spacetime curvature, as happens in $f(R)$ gravity. In such cases, variations of the entropy generate additional terms involving derivatives of $f'(R)$. These contributions prevent the resulting tensor equation from satisfying the Bianchi identity and therefore violate the covariant conservation law $\nabla_\mu T^{\mu\nu}=0$. Hence, as a possible resolution, one can modify the Clausius relation by including an additional entropy-production term $d_iS$ that precisely cancels these spurious contributions such that local conservation of energy is restored. Notably, this correction does not enter into the final form of the field equations. Rather, it acts as a compensating term that ensures that the thermodynamic derivation remains consistent with the Bianchi identity. 

The situation is however conceptually different in the cosmological apparent-horizon (CAH) approach. In this framework the apparent horizon of an FLRW universe is treated as a physical boundary that satisfies a first-law-like thermodynamic relation. The starting point is the Clausius relation applied to the apparent horizon, with the horizon entropy taken to be the Wald entropy and the energy flux across the horizon given by the standard expression. However, for gravity theories with higher-order equations of motion, this equilibrium relation does not generally reproduce the correct Friedmann equations. In particular, when the Wald entropy associated with these modified gravity theories is substituted into the standard Clausius relation, some terms appearing in the dynamical equations derived from the action principle are missing, while additional spurious terms may arise. Consequently, to recover the correct cosmological equations one needs to modify the Clausius relation by introducing an additional entropy-production term so that it takes the form given in Eq.~\eqref{CAH}. The extra contribution $dS_p$ is then interpreted as an entropy-production term whose form is chosen such that the modified Clausius relation reproduces the correct Friedmann equations obtained from the field equations. Since there is no independent thermodynamic or geometric principle that determines the form of this entropy-production term in these theories, its structure must be fixed by comparing with the known cosmological dynamics. As a result, deriving the Friedmann equations from the non-equilibrium Clausius relation requires prior knowledge of the Friedmann equations themselves, hence leading to a certain degree of circularity in the argument.

One must also note that, although the resulting non-equilibrium Clausius relation obtained in this context resembles the one appearing in the EGJ approach, the physical origin of the entropy-production term is fundamentally different. In the EGJ framework, the term $d_iS$ arises directly from the requirement that the equations of motion obtained via the thermodynamic approach remains compatible with the Bianchi identity and satisfies local-matter conservation law. By contrast, in the CAH framework the entropy-production term $dS_p$ is introduced `\emph{a posteriori}' so that the Clausius relation reproduces the exact Friedmann equations, as obtained from the gravitational field equations. Furthermore, since EGJ analysis is independent of choice of the metric, the form of entropy production obtained remains valid in any space-time for a particular gravity theory. However, in CAH framework the form of $dS_p$ term which must be chosen to retain the correct Friedmann equations depends directly on the choice of the space-time metric, in the present case we have chosen non-flat FLRW metric. In other words, in CAH framework there is no general prescription for the entropy production term for a particular gravity theory.

\section{Horizon thermodynamics in modified gravity theories: equilibrium or non-equilibrium ?} \label{eqnoneq}

Alternatively, numerous attempts have been made to reformulate the thermodynamic quantities defined on the apparent horizon, such as the Misner-Sharp mass, horizon entropy density, and the energy flux through the horizon, so that the equilibrium Clausius relation can still be applied in higher-order gravity theories. For example, generalized quasi-local energies have been proposed in which the Misner-Sharp mass is extended to $f(R)$ and scalar-tensor gravity in an FLRW spacetime \cite{Akbar:2006kj}. In such constructions the heat flux through the horizon is correspondingly redefined in terms of this generalized energy. However, these definitions are typically motivated by the gravitational field equations or the Friedmann equations themselves, and therefore the thermodynamic derivation of the dynamical equations may again involve a certain degree of circular reasoning.

A more general approach was proposed in \cite{Gong:2007md}, where it was shown that the correct cosmological equations in modified gravity theories, including $f(R)$ and scalar-tensor gravity, can indeed arise from equilibrium Clausius relation by introducing an appropriately defined masslike function associated with the apparent horizon. However, although such constructions provide an elegant thermodynamic interpretation of the Friedmann equations, all these results collectively illustrate that the definition of thermodynamic quantities on the apparent horizon in modified gravity is not unique.

This exact ambiguity actually reflects a more general feature of horizon thermodynamics in modified gravity theories. In contrast to general relativity, where entropy of the horizon depends only on its area, the Wald entropy in higher-curvature theories can generally depend on additional geometric quantities, such as curvature invariants or scalar fields. As a result, variations of the entropy contain extra contributions involving derivatives of these quantities. Furthermore, since these terms do not admit a unique thermodynamic interpretation, they can be treated as the part of energy flux through the horizon, or equivalently as correction terms to horizon entropy itself, without altering the dynamical equations of the Universe. Hence, depending on how the thermodynamic variables associated with the horizon are defined, these additional contributions may either be interpreted as an internal entropy-production term, leading to a non-equilibrium entropy balance relation of the form $dS=\delta Q/T+d_iS$, or be absorbed into suitably redefined notions of heat flux and quasi-local energy, thereby restoring the equilibrium Clausius relation. In this sense, the same gravitational dynamics can be formulated in either an equilibrium or a non-equilibrium thermodynamic language through an appropriate redefinition of the horizon thermodynamic quantities.

This observation suggests that the distinction between equilibrium and non-equilibrium horizon thermodynamics in higher-order modified gravity theories may not be fundamental, but instead reflecting the freedom in redefining effective thermodynamic variables associated with the horizon. Consequently, the mere appearance of entropy-production terms does not necessarily signal genuine irreversible thermodynamic processes, but may in fact arise from a particular choice of thermodynamic description for the underlying gravitational dynamics.

\section{Conclusion}

In this article, we presented a detailed analysis of the two thermodynamic formulations that relate gravitational dynamics to Clausius relations associated with horizon surfaces: the local Rindler-horizon construction developed by Eling, Guedens, and Jacobson (EGJ), and the cosmological apparent-horizon (CAH) approach used in expanding FLRW spacetimes. Interestingly, we found that, although both frameworks employ exactly similar entropy balance relations that resemble non-equilibrium thermodynamics, the role played by the corresponding entropy-production terms are fundamentally different.

In the EGJ framework the entropy-production term arises when the entropy density of the horizon depends on curvature invariants, as in $f(R)$ gravity. In this case the additional term is required to maintain consistency with the Bianchi identity and the conservation of the energy-momentum tensor. Importantly, this contribution does not enter into the final gravitational field equations, but instead acts as a compensating term that restores the consistency of the thermodynamic derivation.

By contrast, in the CAH framework the entropy-production term is introduced so that the Clausius relation reproduces the Friedmann equations obtained from the gravitational field equations. Since its form must be determined by comparison with the known cosmological dynamics, the resulting thermodynamic derivation may involve a certain degree of circular reasoning. Furthermore, we see that in this case, the additional entropy production term can directly enter into the Friedmann equations hence contributing directly to dynamics of the Universe. This shows that the entropy-production terms appearing in the two approaches are conceptually distinct despite the similarity in their formal expressions.

More generally, our analysis highlights that the thermodynamic description of horizons in modified gravity theories is not unique. The additional contributions arising from curvature-dependent entropy can either be interpreted as internal entropy production or absorbed into redefinition of the heat flux and quasi-local energy associated with the horizon. Consequently, the distinction between equilibrium and non-equilibrium horizon thermodynamics in higher-order gravity theories may largely reflect the freedom in defining effective thermodynamic variables rather than a fundamental physical difference.

Arguments presented in this article raises the fundamental question: Does the apparent non-equilibrium nature of horizon thermodynamics in modified gravity correspond to a genuine physical non-equilibrium effect, or is it merely a consequence of how the thermodynamic variables are defined ?

\appendix

\section{Brief overview of $f(R)$ gravity}\label{fRaction}

The gravitational action for $f(R)$ gravity minimally coupled to perfect-fluid matter takes the form:
\begin{equation}\label{eq:action}
	S = \frac{1}{2\kappa}\int d^4x\,\sqrt{-g} \, f(R) + S_m ,
\end{equation}
where $\kappa\equiv8\pi G$, $R$ is the Ricci scalar and $S_m$ denotes the matter action. Variation of this action integral with respect to the metric, gives the modified field equations;
\begin{equation}\label{eq:feq}
	2f'R_{\mu\nu} - fg_{\mu\nu} + 2f'(g_{\mu\nu}\Box - \nabla_{\mu}\nabla_{\nu})= 2\kappa T_{\mu\nu}
\end{equation}
where, $\Box\equiv g^{\alpha\beta}\nabla_{\alpha}\nabla_{\beta}$. Then, by adopting the FLRW line element given below with spatial curvature term `$k=0,\pm1$', one obtains the Friedmann equations in $f(R)$ gravity.
\begin{equation}
	ds^2 = -dt^2 + a^2(t)\left[\frac{dr^2}{1-kr^2} + r^2 d\Omega^2\right]
\end{equation}
Using the standard definition of the Hubble parameter $H\equiv\dot a/a$, one obtains the following form for Ricci scalar,
\begin{equation}\label{eq:RFLRW}
	R = 6\dot{H} + 12H^2 + 6(k/a^2)
\end{equation}
Assuming the comic component to be a perfect fluid with energy density $\rho(t)$ and pressure $p(t)$, the energy-momentum tensor in co-moving frame takes the form $T^\mu_{\;\nu} = \mathrm{diag}\left(-\rho, p, p, p\right)$. Subsequently, by utilizing the above relations, owe get the Friedmann equations,
\begin{align}
	6f'H^2 &= 2\kappa \rho+ f'R - f - 6H\dot{f'}-6F(k/a^2) \label{eq:friedmann1}\\
	2f'\dot H &= -\kappa(\rho + p) -\ddot{f'} + H\dot{f'}+ 2f'(k/a^2) \label{eq:friedmann2}
\end{align}

\section*{Acknowledgments}
The authors, V. A. P and V. S. N, thank the Cochin University of Science and Technology for providing the financial support in the form of Senior Research Fellowship. During the preparation of this work the authors used ChatGPT (OpenAI) and Grammarly for improving grammar and language usage in the article. After using this tool, the authors reviewed and edited the content as needed and take full responsibility for the content of the published article.

\end{document}